\documentclass[twocolumn]{aastex63}

\usepackage{balance}

\usepackage[T1]{fontenc}

\usepackage[scaled]{helvet}

\usepackage[version=4]{mhchem}

\usepackage{multirow}

\submitjournal{ApJL}

\shortauthors{\textsf{Lichtenberg \& Clement}}

\graphicspath{{./}{figures/}}

\begin{document}

\title{Reduced late bombardment on rocky exoplanets around M-dwarfs}
\shorttitle{\textsf{Reduced late bombardment on M-dwarf exoplanets}}

\correspondingauthor{Tim Lichtenberg}
\email{lichtenberg@astro.rug.nl}

\author[0000-0002-3286-7683]{Tim Lichtenberg}
\affiliation{Atmospheric, Oceanic and Planetary Physics, Department of Physics, University of Oxford, Oxford OX1 3PU, UK}
\affiliation{Kapteyn Astronomical Institute, University of Groningen, PO Box 800, 9700 AV Groningen, NL}

\author[0000-0001-8933-6878]{Matthew S. Clement}
\affiliation{Earth and Planets Laboratory, Carnegie Institution for Science, Washington, DC 20015, USA}
\affiliation{Johns Hopkins APL, 11100 Johns Hopkins Rd, Laurel, MD 20723, USA}

\begin{abstract}
Ocean-vaporizing impacts of chemically reduced planetesimals onto the early Earth have been suggested to catalyse atmospheric production of reduced nitrogen compounds and trigger prebiotic synthesis despite an oxidized lithosphere. While geochemical evidence supports a dry, highly reduced late veneer on Earth, the composition of late-impacting debris around lower-mass stars is subject to variable volatile loss as a result of their hosts' extended pre-main sequence phase. We perform simulations of late-stage planet formation across the M-dwarf mass spectrum to derive upper limits on reducing bombardment epochs in Hadean analog environments. We contrast the Solar System scenario with varying initial volatile distributions due to extended primordial runaway greenhouse phases on protoplanets and desiccation of smaller planetesimals by internal radiogenic heating. We find a decreasing rate of late-accreting reducing impacts with decreasing stellar mass. Young planets around stars $\leq$ 0.4 $M_\odot$ experience no impacts of sufficient mass to generate prebiotically relevant concentrations of reduced atmospheric compounds once their stars have reached the main sequence. For M-dwarf planets to not exceed Earth-like concentrations of volatiles, both planetesimals and larger protoplanets must undergo extensive devolatilization processes and can typically emerge from long-lived magma ocean phases with sufficient atmophile content to outgas secondary atmospheres. Our results suggest that transiently reducing surface conditions on young rocky exoplanets are favoured around FGK- stellar types relative to M-dwarfs.
\end{abstract}


\section{Introduction} \label{sec:intro}

Laboratory simulations of prebiotic synthesis indicate that the favoured geochemical environments for the emergence of life as we know it are highly to moderately reduced \citep{kitadai2018origins,Benner2020,2020SciA....6.3419S}. Key precursor molecules such as hydrogen cyanide (\ce{HCN}), formaldehyde (\ce{CH2O}), cyanamide (\ce{CN2H2}) and cyanoacetylene (\ce{C3HN}) are unstable under modern Earth-like surface conditions. Thus, from a chemical point of view, the prebiotic Earth should have been a very different world than the one we inhabit today.  As a result of Earth's lithosphere and crust composition being close to the quartz-fayalite-magnetite mineral buffer \citep{stagno2021redox}, modern volcanic emissions are dominated by oxidized gases, such as \ce{H2O} and \ce{CO2}. However, geochemical analyses of Archean and Hadean rock samples \citep{2012GeCoA..97...70T,2017Litho.282..316R} suggest similar conditions on the earliest Earth; potentially the result of disproportionate ferrous and ferric iron in the mantle \citep{2019Sci...365..903A,hirschmann2022magma}, as well as hydrogen loss from the early atmosphere \citep{2018A&ARv..26....2L,2020SciA....6.1420C,2021MNRAS.505.2941Y}.

Thus, facing a conundrum between the chemically-favoured environment and its apparent absence in the sampled geochemical record, recent works have turned their attention to the catalytic potential of intermittent reducing atmospheres triggered by iron-rich impactors during the tail-end of Earth's accretion \citep{2003JGRE..108.5070S,2007JGRE..112.5010H,2015Icar..257..290K,2017E&PSL.470...87G,2017E&PSL.480...25G,2017ApJ...843..120S,Benner2020,2020PSJ.....1...11Z,2022PSJ.....3..116C,2022PSJ.....3..115I,2022PSJ.....3...83C}. Impacts of asteroidal and meteoritic materials onto the Earth after the main stage of iron core formation (the `late veneer') have been used to explain the excess abundances of highly siderophile (core-affine) elements in the mantle in near-chondritic abundances \citep[e.g.:][]{2016RvMG...81..161D}. While previously considered to be volatile-rich \citep{2009Natur.461.1227A,2013GeCoA.105..146H}, recent bulk elemental abundances and isotopic evidence from the inner Solar System constrain the composition of this late accretion phase to be volatile-poor and chemically reducing \citep{2016AmMin.101..540H,2017Natur.541..521D,2017Natur.541..525F,2018SSRv..214..121C,2020NatGe..13..265G}.

Phylogenetic evidence \citep{wolfe2018horizontal} and the timing of the earliest robust surface biosignatures constrain the origin of life on Earth to within the first few hundred million years after the Moon-forming impact \citep{2018AsBio..18..343P,Benner2020}, suggesting that life's emergence on other rocky planets might ensue rapidly after habitable surface conditions are attained. However, more than four billion years of processing has left Earth's geological record highly biased toward younger surfaces ages, with the Hadean eon being only accessible via zircon inclusions \citep{2020AREPS..48..291B,2021PreR..359j6178K}.  These limitations make it challenging to definitively determine the timescale for triggering prebiotic synthesis within the young Earth's ambient environment, and thus motivate comparative exploration of extrasolar planets \citep{2021arXiv211204309M,2022arXiv220310023L}.

While G-dwarf systems hosting terrestrial planets would provide an ideal comparison to the Solar System \citep{2019AREPS..47..141J,2021arXiv211204663W}, their scarcity in close proximity to the Sun coupled with the limitations of the transit and radial velocity techniques greatly restrict the detailed exploration of these systems until space-based direct imaging surveys become technically viable \citep{2019AJ....158...83A,2020arXiv200106683G,2021arXiv210107500Q}. However, formation models \citep[e.g.:][]{ogihara09,miguel11} and survey mission yields \citep[e.g.:][]{dressing15,gaidos16} both indicate Earth-sized planets are more commonly hosted in the habitable zones of M-dwarfs.  While the smallest stars offer the best opportunity to study nearby Earth analogs \citep{2007AsBio...7...85S}, their atmospheric and geophysical evolution may follow strongly differing trajectories due to varying volatile delivery pathways \citep{2015NatGe...8..177T,2022arXiv220310056K}, the superluminous pre-main sequence phase and flaring rates of M-dwarfs \citep{2022arXiv220502331M}, and the effects of tidal-locking on climate \citep{2019AnRFM..51..275P}.

The best-characterized M-dwarf system to date, TRAPPIST-1, provides evidence for a break in the planetary bulk composition around the orbital runaway greenhouse threshold \citep{trappist_new}: denser inside (b, c), lighter outside (d and beyond). This can be explained by continuous water loss by photolysis from planets b and c \citep{2015AsBio..15..119L,2018AJ....155..195W,2019A&A...628A..12T,2021AsBio..21.1325B,2021ApJ...922L...4D}. Alternatively, the density constraints may be attributed to incomplete core-mantle segregation \citep{2008ApJ...688..628E} or underdense metal cores \citep{2022PSJ.....3..127S}.
Recent works on late-stage volatile delivery around TRAPPIST-1 and other M-dwarfs suggest that water delivery from the outer disk to short period planets is highly inefficient after nebular dispersal \citep{ray21,Clement22a} in the absence of dynamical perturbations from outer giant planets.  However, admixing and migration processes during the disk phase can result in substantial, $\gtrsim$wt\%-level enrichment of water and other volatile ices \citep{2017A&A...598L...5A,2018NatAs...2..297U,2019A&A...627A.149S,2019NatAs...3..307L,2020A&A...643L...1V,2020SSRv..216...86V,2021ApJ...913L..20L}. Observational evidence for volatile ice-rich (tens of wt\%) super-Earths illustrate the efficacy of volatile admixing into inner planetary systems of low-mass stars \citep{LuquePalle2022Science}.

Like in dynamically unperturbed G-dwarf systems, late accretion around M-dwarfs is dominated by local debris. Therefore, differences in the starting configuration of protoplanets and smaller planetesimals inherited from the early formation phase may induce substantial variations in the timeline and composition of impacts. In this work, we explore the potential for rock planets around M-dwarfs to experience prolonged bombardment of dry, chemically reduced material that may induce observable epochs of transient reducing climates \citep{2019asbi.book..197G,2020AsBio..20.1476F,2020ApJ...888...21R}. Moreover, our models explore the consequences of varied initial conditions inherited from planetary accretion and internal geophysical and geochemical processing during the disk phase across the M-dwarf mass spectrum. We quantify (i) the rate of potentially reduced impacts large enough to trigger intermittent reduced climate states in otherwise oxidised planetary environments, and (ii) the timing of these impacts relative to the extended magma ocean epochs of M-dwarf exoplanets beyond the orbital runaway greenhouse threshold on the stellar main sequence. We outline our methods in Sect.~\ref{sec:methods}, present and interpret our results in Sect.~\ref{sec:results} and \ref{sec:discussion}, and conclude in Sect.~\ref{sec:conclusions}.

\pagebreak
\section{Methods \& Physical Scenarios} \label{sec:methods}
\label{section:methods}

In this section we introduce our choice of numerical methods, starting with a description of the $N$-body simulations of rocky planet formation and early bombardment utilized in our analyses (reported in previous work), followed by an introduction and motivation of the initial conditions (scenarios). A more detailed discussion of the potential of prebiotic synthesis for a given planet formation and evolution scenario is presented in Sect.~\ref{sec:discussion}.

\subsection{N-body simulations and bombardment model}
\label{section:nbody}

\begin{figure*}[tbh]
 	\centering
 	\includegraphics[width=0.95\textwidth]{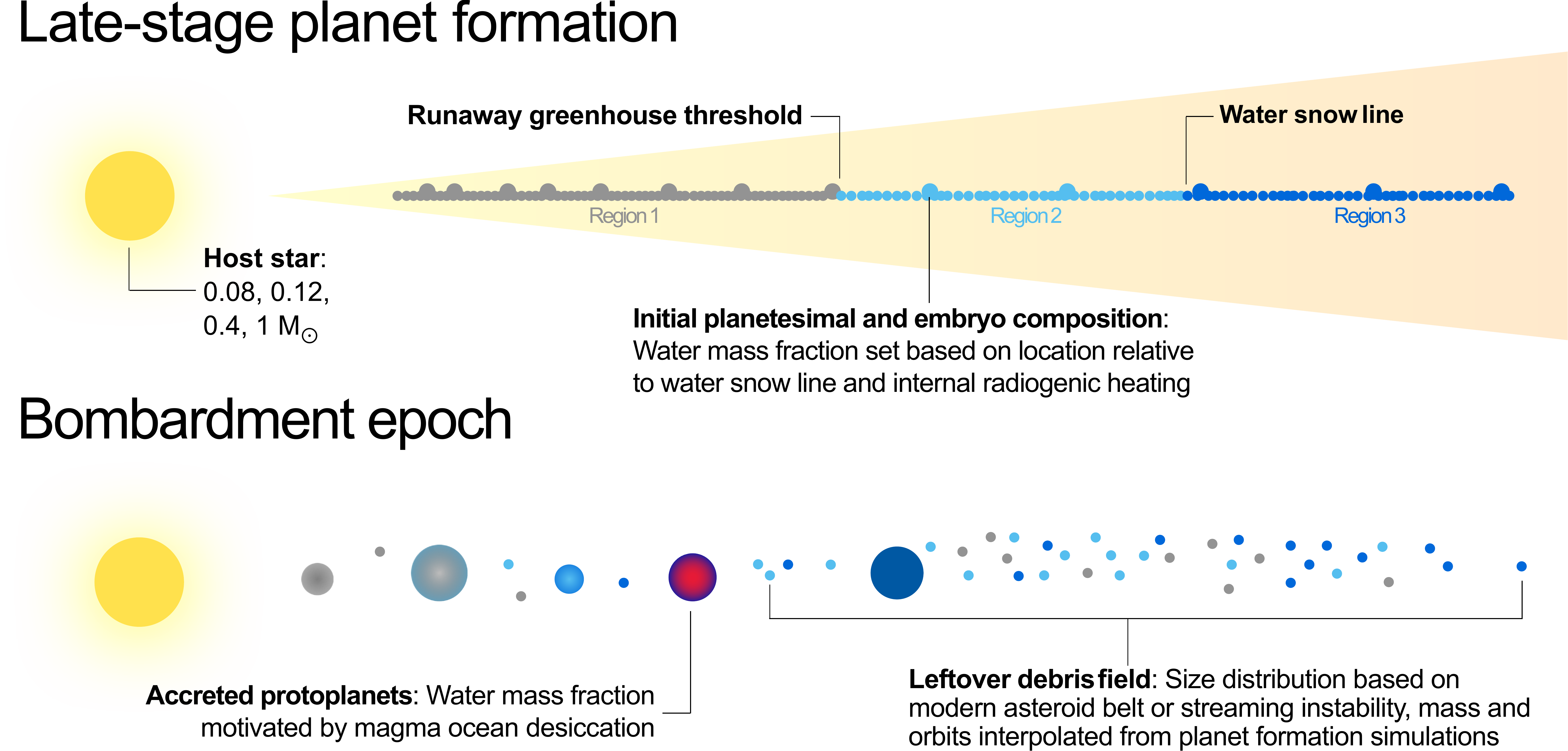}
 	\caption{\textsf{Illustration of the planet formation scenario incorporated in the $N$-body and long-term planetesimal bombardment simulations, described in more detail in \citet[][]{Clement22a}. Varying initial conditions related to embryo and planetesimal water content across the three regions simulate differing physical scenarios of material devolatilization and redox gradients \citep{2022arXiv220310023L} across the planetesimal disk. Scenarios are listed in Tab.~\ref{tab:table1} and discussed in Sect.~\ref{sec:delivery}.}}
     \label{fig:1}
\end{figure*}

Our numerical experiments build on simulations of in-situ planet formation, both in the Solar System and across M-dwarfs with masses 0.08 to 0.6 $M_{\odot}$, as reported in \citet{clement21_tp,Clement22a}. All presented $N$-body simulations use the \textsc{Mercury6} hybrid integrator \citep[][]{chambers99} and standard settings that are commonly utilized in $N$-body studies of planet formation \citep[e.g.: surface density profile, disk mass, orbital eccentricities and inclinations of the particles, and integration time-step as a fraction of the innermost particle's orbital period: ][]{chambers01,ray07_mdwarf,ray09a}.  Other parameters such as the inner and outer radii of the terrestrial disk are necessarily scaled by stellar mass.  Each of 216 total M-dwarf simulations consider a disk of 20 planetary embryos and 400 planetesimals that extends from 0.01--0.5 au (top panel of Fig.~\ref{fig:1}).  The data presented in our current investigation combines simulation outputs from models considering total disk masses of 3.0 and 6.0 $M_{\oplus}$, with the solid density decreasing with radius as $\propto r^{-3/2}$.  We compare the M-dwarf models with reference models of the formation of the Solar System's terrestrial planets that start with a 5.0 $M_{\oplus}$ disk of 100 embryos and 1,000 planetesimals, which span the orbital range from 0.5--4.0 au.  In order to make the most accurate comparison with the Solar System possible, we choose 216 reference simulations that incorporate perturbations from the evolving giant planets \citep[the Nice Model instability,][]{nesvorny12} that have previously been tested and validated against a number of important dynamical and cosmochemical constraints \citep{deienno18,clement19_ab,mojzsis19,nesvorny21_tp}.

We derive hypothetical delayed bombardment chronologies for systems with detected exoplanets in the habitable zones of host-stars with masses of 0.08, 0.12, 0.4, and 1.0 $M_\odot$.  We first select all remaining planetesimals and collisional fragments from the planet formation simulations described in \citet{clement19_frag} and \citet[][]{Clement22a}, which includes an imperfect accretion algorithm to generate fragments.  For our M-dwarf models, we combine 1,375 surviving planetesimals from the original planet formation simulations with a population of 1,285 collisional fragments generated by performing $\sim$10$^{5}$ follow-on simulations of the final giant impacts in each planet formation model.   We then integrate the orbital evolution of each of these massless particles (a size-frequency distribution is added when post-processing the simulation-derived bombardment chronologies)  in several planetary systems.  These include: TRAPPIST-1 \citep{gillon17,trappist_new}, Proxima Centauri \citep{proxima_new,proxima_c}, TOI-700 \citep{gilbert20,rodriguez20}, and GJ 229 \citep{feng20a}.  To ensure the debris field spans the appropriate radial range of each system, we interpolate the semi-major axes of each particle to appropriately scale our planetesimal and fragment populations to each system of interest.

For each system we then derive bombardment curves for the entire system, and for planetesimals and fragments originating from several radial bins at the beginning of the simulation via exponential fitting of the simulation impact chronologies. We combine these functions with the total masses of leftover material and a size frequency distribution (SFD) of the debris field to generate late impact histories for the different model systems. For all model scenarios presented in this work, we set the total mass of leftover material to 0.1 $M_{\oplus}$, a value that is approximately equivalent to the average total mass in remaining particles at the end of our planet formation simulations, and is consistent with the geochemically inferred mass of the late veneer \citep[e.g.,][]{raymond13}. To account for potentially different dominant accretion processes during the gas disk phase, we consider two different debris SFDs: one that is set to mimic that of the modern asteroid belt down to 1.0 km sizes, and a second analogous distribution that incorporates an additional $\sim$ 70--120 km component (20$\%$ of the total debris mass) that is motivated by the results of planetesimal formation simulations \citep{johansen15,2019ApJ...885...69L} and the sizes of main belt asteroids not belonging to collisional families \citep{delbo17}.

It is worth discussing how the resolution of our models might affect our results.  While modern integration algorithms \citep[e.g.:][]{grimm14,rein15} are capable of modeling the process of terrestrial planet formation with many more particles than utilized in our work, studies varying the number of embryos and planetesimals used have noted only minor differences in the distributions of the final systems' dynamical architectures \citep{clement20,woo21}.  Thus, as the purpose of our modeling effort is to generate debris populations and study the rate of reduced impacts on potentially habitable planets, we opted to save compute time and co-add the leftover planetesimal and fragment populations from a large number of initial, lower-resolution planet formation simulations.  Similarly, it is also now possible to study late bombardment more directly with high-resolution simulations. \citet{Clement22a} compared the statistical methodology described above with the results of high-resolution, GPU-accelerated bombardment simulations directly incorporating our model SFD and found the methodologies produced qualitatively similar results.

\subsection{Planetesimal composition and impact timing}
\label{section:wmf}

\begin{table*}[tbh]
    \begin{center}
    \begin{tabular}{| c | c || c | c | c | c | c | c |}
    \hline
    \multicolumn{2}{|c||}{\multirow{2}{*}{Scenario}} & \multicolumn{2}{c |}{WMF \emph{Region 1}} & \multicolumn{2}{c |}{WMF \emph{Region 2}} & \multicolumn{2}{c |}{WMF \emph{Region 3}}\\
    \multicolumn{2}{|c||}{} & Embryos & Planetesimals & Embryos & Planetesimals & Embryos & Planetesimals \\
    \hline \hline
    \multicolumn{2}{|c||}{\emph{Solar System Reference}} & 0.001 & 0.001 & 0.001 & 0.001 & 0.1 & 0.1 \\
    \hline  \hline
    A & \emph{Icy 1}  & 0.001 & 0.001 & 0.001 & 0.001 & 0.25 & 0.25 \\
    B & \emph{Icy 2}  & 0.001 & 0.25 & 0.001 & 0.25 & 0.25 & 0.25 \\
    \hline
    C & \emph{Dry 1} & 0.0 & 0.001 & 0.0 & 0.01 & 0.0 & 0.1 \\
    D & \emph{Dry 2} & 0.0 & 0.001 & 0.0 & 0.001 & 0.1 & 0.1 \\
    E & \emph{Dry 3} & 0.0 & 0.0 & 0.0 & 0.0 & 0.1 & 0.01 \\
    \hline
    F & \emph{Desiccated Planetesimals 1}  & 0.001 & 0.01 & 0.001 & 0.01 & 0.1 & 0.01 \\
    G & \emph{Desiccated Planetesimals 2}  & 0.001 & 0.0 & 0.001 & 0.0 & 0.1 & 0.01 \\
    H & \emph{Desiccated Planetesimals 3}  & 0.001 & 0.0 & 0.001 & 0.0 & 0.1 & 0.001 \\
    \hline
    I & \emph{Desiccated Embryos 1} & 0.0 & 0.01 & 0.01 & 0.25 & 0.1 & 0.25 \\
    J & \emph{Desiccated Embryos 2} & 0.0 & 0.01 & 0.0 & 0.1 & 0.0 & 0.25 \\
    \hline
    & \multicolumn{1}{r ||}{$t=0$ Myr} & 0.001 & 0.001 & 0.001 & 0.001 & 0.1 & 0.1 \\
    K & \multicolumn{1}{r ||}{\emph{Evolve embryos} \hspace{0.1cm} $t=1$ Myr} & 0.0 & 0.001 & 0.001 & 0.001 & 0.1 & 0.1 \\
    & \multicolumn{1}{r ||}{$t=10$ Myr} & 0.0 & 0.001 & 0.0 & 0.001 & 0.1 & 0.1 \\    
    \hline
    \end{tabular}
    \end{center}
    \caption{\textsf{Initial water mass fraction (WMF) distributions of planetesimals and planetary embryos used in the analyses plotted in Figure \ref{fig:3}.  The first column reports the model designation, and the subsequent six columns provide the initial WMF of embryos and planetesimals in \emph{Region 1, 2} and \emph{3} (see Fig.~\ref{fig:1}).  Note that, in the \emph{Evolve Embryos} scenario, the disk is initialized in the same manner as in the reference case, however after 1 Myr of simulation time the water mass fractions of all embryos in \emph{Region 1} (including growing planets) changes to 0$\%$, and \emph{Region 2} embryos desiccate at $t=$ 10 Myr. The physical motivation for each model scenario is discussed in Sect.~\ref{sec:delivery}.}}
    \label{tab:table1}
\end{table*}

We assess the plausible spectrum of final water mass fractions (WMF) and bombardment histories of our models' fully formed planets in the liquid-water habitable zone via post processing of the planets' accretion histories. Figure \ref{fig:1} illustrates the compositional assumptions and Table \ref{tab:table1} summarizes the assumptions of 11 different disk compositional gradients that might result from disparate nebular disk thermal states and gradients during the formation \citep{2022arXiv220309759D,2022arXiv220310056K} and internal evolutionary processes of planetesimals and protoplanets \citep{2022arXiv220310023L}.  Each model divides the disk into three different regions (Fig.~\ref{fig:1}). The precise number of initial particles in each region depends on the stellar mass model used, and varies between $\sim$4--10 embryos and $\sim$60--185 planetesimals.  In general, models utilizing a lower stellar mass possess larger numbers of particles initially in the outermost radial bins.  The first, and usually driest, region extends from the inner edge of the distribution of planet forming material to the interior boundary of the conservative habitable zone (\emph{Region 1}). We determine the inner boundary of this second region (\emph{Region 2}) following \citet{hz}, defined by the steam runaway greenhouse limit.  This second orbital regime  encompasses all of the conventional liquid water habitable zone, and stretches out to the location of the water ice line during the evolution of the protoplanetary disk.  Planetesimals and embryos in the remaining section of the disk (\emph{Region 3}) are usually assigned the largest WMF values in the majority of our models.  This assumes that water-rich planetesimals can easily form in \emph{Region 3}, and that those objects are typically processed the least from internal radioactive heating because of time delays in planetesimal formation rates at larger orbital separations \citep{2022arXiv220309759D}.

The redox state of planetesimals and planetary embryos, usually measured by the FeO concentration in a sample, is crucially affected by the abundance of water, which can raise the overall oxidation state of planetary materials by reaction with other compounds \citep{2008ApJ...688..628E,2008ApJ...685.1237E,2022arXiv220310023L}. The increased oxidation state of magmatic iron meteorites \citep{2018NatGe..11..401B,2022GeCoA.318..112H}, evidence for degassing from their parent planetesimals \citep{2021Sci...371..365L,2021PNAS..11826779H}, fluid flow \citep{2016M&PS...51.1886L} and hydrogen incorporation \citep{2020Sci...369.1110P,2021PSJ.....2..244J} in ordinary and enstatite chondrites, accretion of water onto achondrites \citep{2017RSPTA.37560209S} and the devolatilization trend in planetary materials \citep{2019Icar..328..287W,2019MNRAS.482.2222W}, both in carbonaceous and non-carbonaceous meteorites \citep{Alexander2019a,Alexander2019b}, all suggest a substantial initial abundance of highly volatile elements (H, C, N) in inner planetary systems during the disk phase. In addition, uncertainties in high-pressure metal-silicate partitioning allow the earliest accretion phase to be dominated by oxidized materials, fulfilling present-day constraints on the Earth's mantle composition by a transition from oxidized to reduced \citep{2013Sci...339.1194S,2020PNAS..11727893H}.  This contrasts the traditional assumption that material composition transitions from reduced to oxidized during planetary accretion \citep{2015Icar..248...89R,2018SSRv..214...47O}. Here, we use the water mass fraction before, during, and after planetesimal and protoplanet evolution as the primary marker for an object's redox state and material composition.

Following \citet{2020PSJ.....1...11Z} and \citet{2022PSJ.....3..115I}, we are primarily interested in planetesimal impacts in the size range of $\sim$200--1000 km. Smaller impacts will have a substantially shorter time span of inducing transiently reducing conditions because of their limited ability to vaporize liquid water oceans. Impacts too large, on the other hand, would melt the planetary mantle; generating a surface magma ocean and burying the planetesimal iron core in the mantle without chemically equilibrating the metal with vaporized water \citep{2022PSJ.....3..116C}. The atmosphere-reducing efficacy of late-accreting debris is further dependent on the timing of the impact event. M-dwarf planetary systems undergo an extended luminous pre-main sequence phase, during which initially water-rich or hydrogen-dominated planets undergo prolonged magma ocean phases \citep{2016ApJ...829...63S,2021AsBio..21.1325B,2021arXiv211204663W,2021JGRE..12606711L,2022arXiv220310023L}. Prebiotically relevant impact events onto formed planets must thus happen after the runaway greenhouse transition has receded to shorter orbital distances. In addition to scrutinizing the timing of the largest events, we also calculate the total amount of water addition from the left-over debris field. Through this process, we are able to derive limits beyond which surface water excesses or deficiencies of several orders of magnitude during the impact event would exclude prebiotically-relevant atmospheric chemistry.

\section{Results} \label{sec:results}

We describe the results of our simulations here divided into two categories. In Sect.~\ref{sec:timing} we focus solely on the timing and intervals of bombardment episodes, while in Sect.~\ref{sec:delivery} we present the influence of the different chemical scenarios on the composition of the late-stage bombardment and resulting planets.

\subsection{Timing of impacts} \label{sec:timing}
\begin{figure*}[htb]
 	\centering
 	\includegraphics[width=0.85\textwidth]{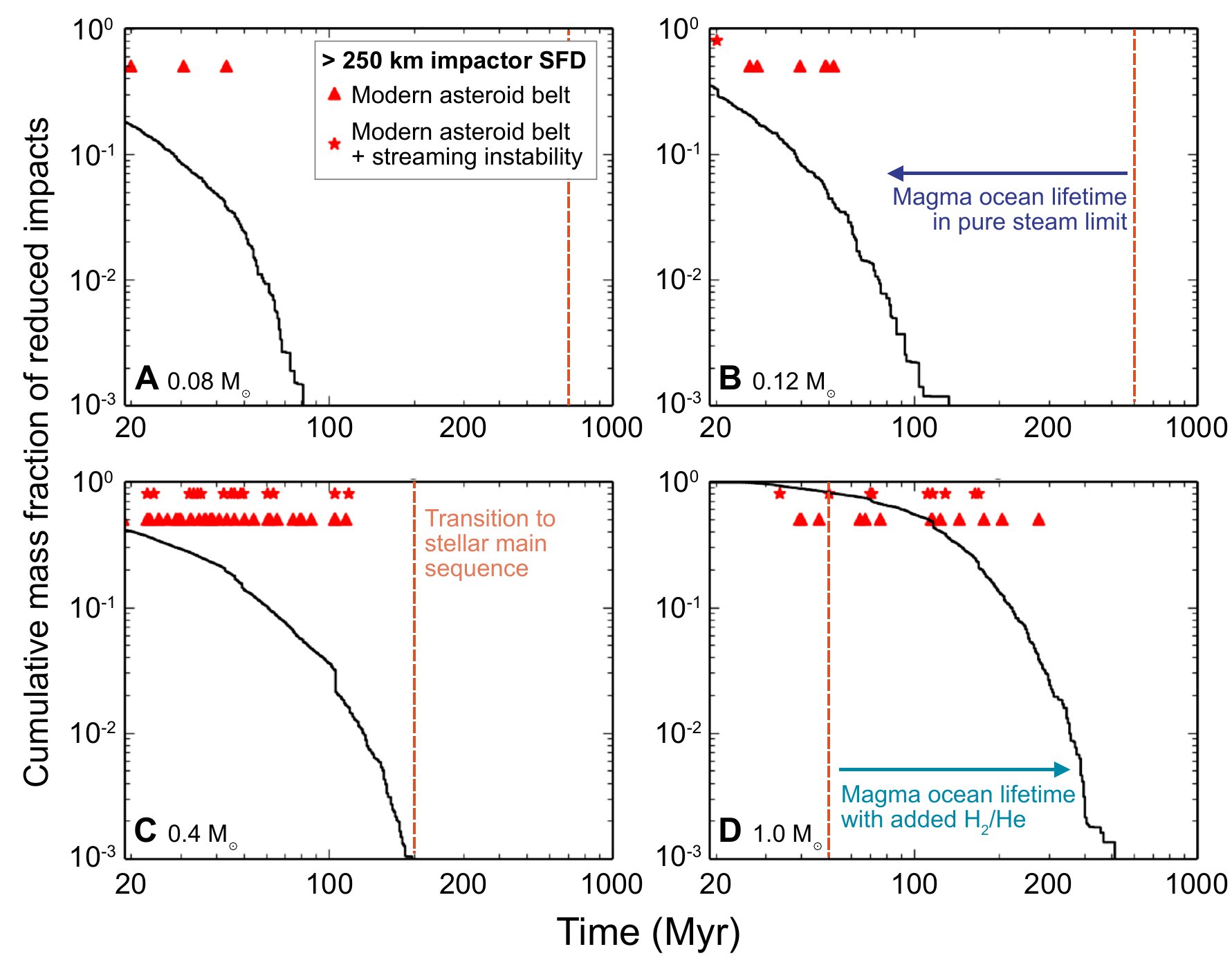}
 	\caption{\textsf{Timing of reduced impacts onto potentially habitable exoplanets around stars of variable mass (panels \textbf{A} to \textbf{D}). The black lines illustrate the (inverse) cumulative debris mass fraction over time sourced from inside the water snow line. Red symbols indicate the timing of impact events larger than 250 km in diameter, an approximate efficacy threshold to trigger intermittent reduced climate states. Triangle symbols represent cases where the left-over debris field is sourced from a size frequency distribution (SFD) comparable to the modern asteroid belt.  Star symbols include an additional 5\% of 80--130 km objects, similar to the peak of the birth planetesimal population generated by the streaming instability. Orange vertical lines demarcate the arrival of the star in each scenario onto the main sequence. In the pure steam limit (water-dominated atmospheres) the magma ocean lifetime can be shorter than this (blue arrow in \textbf{B}), in scenarios with substantial primordial H$_2$/He atmospheres (green arrow in \textbf{D}) the magma ocean lifetime is enhanced. Only for G-dwarfs, such as the Sun, appreciable quantities of late bombardment hit potentially habitable planets beyond the pre-main sequence phase.}}
    \label{fig:2}
\end{figure*}
Fig.~\ref{fig:2} illustrates the timing of arrival of late-impacting debris onto planets beyond the runaway greenhouse transition, in potentially habitable orbits, after the initial magma ocean phase. For cases 0.08 $M_\odot$ (panel \textbf{A}), 0.12 $M_\odot$ (panel  \textbf{B}), and 0.4 $M_\odot$ (panel \textbf{C}), bombardment is focused on the pre-main sequence phase \citep[orange-dashed transition timings from][]{2015A&A...577A..42B}. Specifically for lower-mass M-dwarfs, late impacts cease at around 100 Myr. Only a handful of impacts are massive enough to potentially induce prebiotically relevant reduced climate states. For 0.4 $M_\odot$ (panel \textbf{C}), the end of the bombardment epoch coincides approximately with the transition to the main sequence and the amount of larger impacts increases substantially. However, such impacts predominately occur in the first $\sim$100--150 Myr after system formation. For Solar-type stars, 1.0 $M_\odot$ (panel \textbf{D}), the bombardment epoch extends to about 300--400 Myr while the main sequence-transition is shifted to much earlier times ($\approx$50 Myr). This means that for G-dwarfs more than 80\% of debris arrives during the stellar main sequence. Impacts of large objects for G stars continue until about 200--300 Myr after system formation.

\subsection{Volatile delivery across physical scenarios} \label{sec:delivery}
\begin{figure*}[htb]
 	\centering
 	\includegraphics[width=0.95\textwidth]{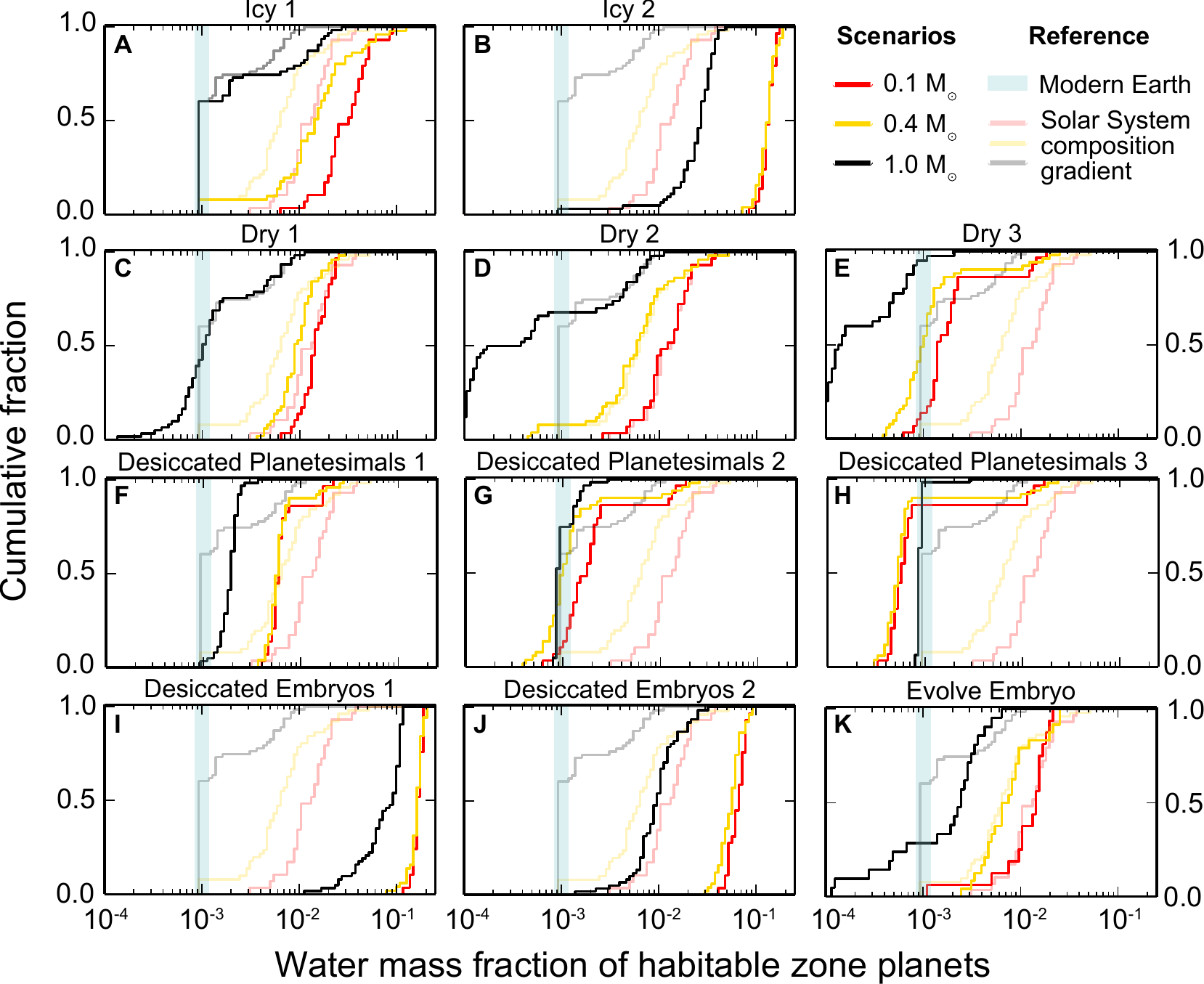}
 	\caption{\textsf{Water mass fraction distribution across the model scenarios (panels \textbf{A} to \textbf{K}) in Tab.~\ref{tab:table1} for 0.1 (red), 0.4 (yellow) and 1.0 (black) $M_\odot$ stars with the fiducial Solar System composition gradient (light grey, yellow and red lines) as reference cases. The modern Earth water mass fraction with uncertainties is indicated as a light blue-green band. The physical motivation for each model scenario is discussed in Sect.~\ref{sec:delivery}. Across most scenarios, M-dwarf exoplanets receive orders of magnitude more water than the modern Earth. Earth-like WMFs for M-dwarf exoplanets are produced with $>$50\% probability for scenarios \emph{Dry 3} (panel \textbf{E}) and \emph{Desiccated Planetesimals 3} (panel \textbf{H}).}}
    \label{fig:3}
\end{figure*}
Fig.~\ref{fig:3} shows our results for the delivery of water -- and hence oxidizing power --  during the bombardment phase of M- and G-dwarf planetary systems. We sub-divide the results into various classes of scenarios: a \emph{Solar System Reference} case, two end-member scenarios of initially very volatile-rich (\emph{Icy}) and volatile-poor (\emph{Dry}) systems, and intermediate scenarios that account for radioactive heating of small planetesimals (\emph{Desiccated Planetesimals}) and magma ocean desiccation (\emph{Desiccated Embryos}) in isolation, while (\emph{Evolve Embryos}) additionally accounts for the time sequence of volatile loss by photolysis from molten protoplanets.  We discuss the physical motivation for each model and the results of the respective bombardment simulations below.

\paragraph{Solar System Reference} 

The Solar System Reference scenario illustrates the dynamical and compositional case of the Solar System after the gas disk phase: relatively dry, volatile-poor (0.001 wt\%) bodies inside the asteroid belt \citep{2019SciA....5.8106J,Alexander2019a,2020Sci...369.1110P,2021PSJ.....2..244J,2021GeCoA.297..203S}, and somewhat wetter but volatile-depleted (0.1 wt\%) outside. Carbonaceous asteroids hosting a few wt\% of water on average \citep{2018SSRv..214...36A,Alexander2019b} motivate our outer compositions. These inferred bulk abundances of water and other volatile ices \citep{2022arXiv220310056K,2022AGUA....300568K} relative to the primordial, disk-inherited compositions (closer to that of comets: $\gtrsim$25 wt\%) represent the combined influence of internal radiogenic heating \citep{2017SciA....3E2514B,2021Sci...371..365L} and collisional overprinting. The dynamical configuration of small bodies in the Solar System was also sculpted by stochastic scattering events with the giant planets, which transports a significant amount of volatile-rich debris toward the inner Solar System, enriching the terrestrial planets during the end-phase of planetary accretion \citep{2017Icar..297..134R,deienno22}. In the Solar System Reference scenario (Fig.~\ref{fig:3}, grey line), about 60\% of planets accrete Earth-like abundances of volatiles (the local baseline), with the upper $\approx$30--40\% still within the upper limit of the Earth's water bulk abundance \citep{2017SSRv..212..743P}.

\paragraph{Icy (A/B)} 
The \emph{Icy} scenarios assume that -- unlike in the Solar System -- internal heating from short-lived radioactive elements like $^{26}$Al is significantly reduced and that magma ocean desiccation during the pre-main sequence phase does not operate. Volatile ice abundances in planetesimals are thus set by the local disk chemistry \citep{2021PhR...893....1O,2022arXiv220309818M}. \emph{Icy 1} assumes that internal radioactive heating in planetesimals was comparable in \emph{Region 1} and \emph{Region 2}, but non-existent beyond the snowline, either due to inhomogeneous enrichment \citep[higher abundances of $^{26}$Al in the inner disk,][]{2021ApJ...919...10A} or slower planetesimal formation and hence reduced heating in the outer disk \citep{2018A&A...614A..62D,2021A&A...652A..35C}. In this situation planets in the 1.0 $M_\odot$ simulation accrete somewhat more water from outside-in scattering of \emph{Region 3} planetesimals. In contrast, the M-dwarf simulations accrete significantly more water, between 0.01 and 0.1 wt, one to two orders of magnitudes above Earth's value. In \emph{Icy 2} planetesimals in \emph{Region 1/2} are also ice-rich. This corresponds to a scenario where planetesimal formation generally occurs at the outward- (during the Class I stage) and then inward-moving (Class II stage) density jump across the water snow line \citep{2017A&A...608A..92D,2017A&A...602A..21S} while the disks forms, and in the absence of radioactive heating. In this case the M-dwarf simulations sample near-maximum abundances of water throughout, between 0.1--0.2 wt, while the G-dwarf planets reach between 0.01--0.1 wt on average.

\paragraph{Dry (C/D/E)} 
The \emph{Dry} scenarios assume that both desiccation by magma ocean losses and internal radiogenic heating of planetesimals are operating to various degrees of efficacy. In \emph{Dry 1}, protoplanets completely desiccate from magma ocean losses and planetesimals dehydrate with decreasing efficacy from \emph{Region 1} to \emph{Region 3} due to slower planetesimal formation with increasing distance (or, alternatively, higher $^{26}$Al in the inner disk). In \emph{Dry 2}, planetesimal dehydration in \emph{Region 2} is enhanced from 0.01 to 0.001 relative to \emph{Dry 1}. In both scenarios, M-dwarf planets receive on average about 0.01 wt water, while G-dwarf planets are drier: about 50\% above Earth levels for \emph{Dry 1} and about 40\% above Earth for \emph{Dry 2}. However, \emph{Dry 2} forms much drier planets on average, down to about 0.0001 wt planets for the most volatile-depleted simulations. The \emph{Dry 3} setting increases planetesimal dehydration further, such that G-dwarf planets barely reach Earth-like water levels, while M-dwarf planets now accrete to approximately Earth-like levels through late-stage debris.

\paragraph{Desiccated Planetesimals (F/G/H)} 
This scenario fixes magma ocean losses and increases planetesimal dehydration via radioactive heating by region (inside-out) and total magnitude. For the G-star simulations the effects from changes in the initial planetesimal water mass fractions are minor across our range of tested scenarios, such that the difference between \emph{Desiccated Planetesimals 1} and \emph{Desiccated Planetesimals 3} result in a statistical dehydration of rocky planets between about 2--3 times. For the M-star simulations, however, the increasing dehydration from \emph{Desiccated Planetesimals 1} to \emph{Desiccated Planetesimals 3} is stark, resulting in a shift from about 0.01 wt water levels for \emph{Desiccated Planetesimals 1} to about 80\% of planets being drier than Earth for \emph{Desiccated Planetesimals 3}. This illustrates the strong effect of early volatile loss by internal radioactive heating of planetesimals during the late-stage bombardment episode of rocky planet formation.

\paragraph{Desiccated Embryos (I/J)} 
These scenario setups pose the general question of whether magma ocean water losses can outcompete planetesimal dehydration to achieve Earth-similar water mass fractions for rocky planets in potentially habitable orbits by increasing the loss levels for embryos. In \emph{Desiccated Embryos 1}, embryo water mass fractions decrease from 0.1 wt in \emph{Region 3} to 0.0 wt in \emph{Region 1}. This assumes that the time scale the planets spend inside the orbital runaway greenhouse threshold is a major factor in desiccation efficacy \citep{2016ApJ...829...63S,2018AJ....155..195W}, and that the small planetesimals in \emph{Region 2} and \emph{Region 3} are primitive, icy, and unaffected by radioactive heating. These simulations produce water mass fractions on the order of 0.1 wt; about 2--3 orders of magnitude above Earth's levels. In \emph{Desiccated Embryos 2} all embryos are completely dried out from magma ocean loss effects, and the planetesimals' water mass fraction in \emph{Region 2} is reduced. In this situation G star planets achieve on the order of 0.01 wt and M star planets receive on the order of 0.05--0.1 wt of water; one to two orders of magnitude above Earth's water mass fraction.

\paragraph{Evolve Embryo (K)}
This scenario assumes that planetesimal internal heating is comparable to the Solar System and magma ocean losses are time-dependent for the first 10 Myr after system formation, when pre-main sequence stars peak in luminosity. In this scenario, about 25\% of the G star simulations become drier than Earth, while 75\% end up being up to one order of magnitude more water rich ($\leq$0.01 wt). M star simulations reach about 0.005--0.03 wt on average; an order of magnitude above Earth's water levels.

\section{Discussion} \label{sec:discussion}

\subsection{Impact-generated reduced climate states on exoplanets}

Our bombardment simulations suggest that the late-stage delivery of reducing power via iron-rich impacts in M-dwarf systems is (i) statistically rare and (ii) likely not capable of triggering intermittent reducing climates at all. Because the left-over debris mass in M-dwarf systems after the main formation era is already heavily depleted compared to G star systems, late-stage impacts are focused on earlier times (Fig.~\ref{fig:2}). Late-stage impacts acting as a \emph{Deus Ex Machina} for sub-aerial prebiotic chemistry in M-dwarf systems are potentially problematic on two accounts. First, the impact flux peaks at very early times, and is likely too early for impacts to play a major role in resetting atmospheric chemistry. If an iron-rich impactor strikes an already reduced atmosphere, the timescale of opportunity \citep{2020plas.book....3Z} is not simply multiplied because atmospheric reduction in the scenario is dependent on the presence of a surface ocean, which is related to the total mass budget of liquid water at the surface. Second, the arrival of the most massive impactors, which are the most effective in triggering extended reduced atmospheres by equilibrating with sufficient surface water, is limited to a few tens of Myr in M star systems. At this time the runaway greenhouse threshold is far outside the orbit of planets that reside in potentially habitable regions during the stellar main-sequence phase \citep{2015AsBio..15..119L,2016ApJ...829...63S}; meaning they will be in a global magma ocean regime \citep{2021arXiv211204663W,2022arXiv220310023L}, because a fraction of an Earth's ocean mass is already sufficient to keep the planetary surface temperature above the melting temperature of rocks \citep{2021ApJ...919..130B}. If the planetary mantle is a magma ocean, impactor iron merges with the target body core \citep{2016E&PSL.448...24K}.

This outcome is focused on M-dwarfs due to their luminous pre-main sequence phase. G-dwarf planets, on the other hand, experience numerous large impacts over an extended time period of hundreds of Myr. If the Solar System is typical in its inner-system composition then G-dwarf exoplanets may undergo similar bombardment episodes during the tail-end of accretion. However, as our simulations across different compositional setups (Fig.~\ref{fig:3}) demonstrate, this conclusion is sensitively dependent on the mechanism of dehydration. 

While recent work suggests that even G-dwarf exoplanets can undergo major shifts in their atmospheric composition during their magma ocean epoch \citep{2018AJ....155..195W,2021AGUA....200294K}, unlike planets inside the runaway greenhouse threshold \citep{2015ApJ...806..216H,2016SSRv..205..153M,2018SSRv..214...76I}, G-dwarf planets in the liquid water habitable zones of their stars will not desiccate to completion.  Because rocky planets in this case can solidify efficiently, ocean formation timescales \citep{2017JGRE..122.1458S} are smaller than the average time interval between subsequent large impactors, thus maximizing the influence of each individual impactor per unit mass. The size frequency distribution of impacting debris contributes to this conclusion. A distribution similar to the modern asteroid belt achieves massive impacts throughout all planetary systems, including the M-dwarf mass spectrum down to 0.1 $M_\odot$. In this mass domain, no massive impacts at all are generated if the SFD incorporates an increasing fraction of birth planetesimals formed through the streaming instability (cf. Fig.~\ref{fig:2}). This suggests that collisional and secondary processing during planetary accretion tend to extend the timescale for late-accreting debris \citep{morby09,2016ApJ...821..126Q}.

So far we have solely analysed the timing of impacts, but what about their composition? For impacts to catalyse the atmospheric production of reduced nitrogen compounds, the impactor itself has to be reduced. In the Solar System only few known meteorites (for instance, enstatite chondrites or aubrite achondrites) are expected to feature compositions that would allow this. All other known meteorite classes -- even magmatic irons and H chondrites -- overlap with carbonaceous chondrites in their whole-body oxidation states \citep{2018NatGe..11..401B,2022GeCoA.318..112H,Corrigan2022}. This contrasts with the standard assumption in the literature that the inner and outer Solar System compositions are solely caused by to the location of the water snowline during the disk phase \citep{2018SSRv..214...47O}, motivating alternative redox and thermal trajectories for the inner Solar System planetesimal population \citep{1993Sci...259..653G,2013Sci...339.1194S,Alexander2019a,2020PNAS..11727893H,2021Sci...371..365L}. The uncertainty on the earliest geophysical evolution of the inner Solar System planetesimal population is underlined by recent evidence for fractionation of refractory and moderately volatile elements \citep{2017Natur.549..511H,2017Natur.549..507N}, which suggests that the terrestrial planet-forming planetesimals experienced significant melting and vaporization at least partially caused by internal heating from short-lived radioactive isotopes \citep{2019Icar..323....1Y,2020Icar..34713772B}. If the present-day composition of inner Solar System planetesimals is inherited from prior redox evolution, the consequences for the atmospheric diversity among the rocky exoplanet population may be profound \citep{2019NatAs...3..307L,2021ApJ...913L..20L,2022arXiv220310056K}; as further illustrated by our late-stage bombardment simulations. Taking the water mass fraction of planetesimals as a proxy for whole-body oxidation state, only simulations that substantially dehydrate from internal radioactive heating experience any late-stage impacts of reduced debris. This suggests that reduced climates on rocky exoplanets around M-dwarfs are unlikely to be caused by bombardment. Therefore, internal geophysical and geochemical processes to escape planetary self-oxidation \citep{2021SSRv..217...22G}, such as rain-out quenched magma ocean regimes \citep{2021ApJ...914L...4L}, may instead enable long-lived outgassed atmospheres rich in reduced compounds \citep{2020E&PSL.55016546L,2021arXiv211105161L}. Alternatively, HCN \citep{Todd2020} or prebiotic organics \citep{Paschek2021,Paschek2022} may be synthesized inside the planetesimals and delivered directly with the impact. 

\subsection{Late-stage volatile delivery in M-dwarf systems}

While aimed at providing insight into the likelihood of reduced impacts on rocky exoplanets, our simulations also provide a different viewing angle on the volatile delivery around smaller stars. While late-stage impact bombardment in the Solar System previously had been assumed to be volatile-rich, the higly siderophile element concentrations in Earth's lithosphere require the so-called late veneer to be dominated by dry, reduced material \citep{2016RvMG...81..161D,2017Natur.541..521D,2017AREPS..45..389K,2018SSRv..214..121C}. Given the results of our simulations and the accompanying discussion in the previous subsection, this conclusion directly translates into consequences for the volatile abundance in extrasolar planets. All but the driest of bombardment simulations yield a significant (up to several orders of magnitude) overabundance of water relative to the terrestrial planet population. In the context of sub-Neptunes, \citet{2021JGRE..12606639B} argued that dry planets can be formed by protoplanets accreting locally and growing from an influx of abundant dry pebbles inside the water snow line (their "drift" end-member scenario). This would provide a physical mechanism for the currently dominant explanation for the Kepler radius valley \citep{2017AJ....154..109F}: escape of primordial \ce{H2}/He from otherwise volatile-poor cores \citep{2017ApJ...847...29O,2018MNRAS.476..759G,2021MNRAS.503.1526R}. However, models invoking the migration of protoplanets \citep{2020A&A...643L...1V,2020SSRv..216...86V} and of the water snow line \citep{2000ApJ...528..995S,2018A&A...614A..62D} during the disk phase suggest otherwise, for which increasing observational evidence around M-dwarf stars is found \citep{LuquePalle2022Science}.

Our simulation results indicate that, even if protoplanets were to escape the main accretion era dominantly dry, they would receive an abundant influx of late-accreting volatile-rich debris. This influx is high enough to overcompensate even the largest atmospheric escape fluxes predicted for ultrashort-period exoplanets. For example, \citet{2018A&A...619A...1B} estimated the maximum water mass to escape from 55 Cancri -- a highly irradiated super-Earth -- to be $\approx$100 Earth oceans; and more likely in the range of a few to a few tens of Earth oceans. Similarly, \citet{2015AsBio..15..119L} and \citet{2018AJ....155..195W} find total escape fluxes to be limited to a few tens of Earth oceans for a wide range of irradiation environments and planetary masses. In our simulations only the \emph{Desiccated Planetesimals} scenarios achieve water mass fractions low enough for atmospheric escape to strip the planets completely of their water contents. This is because high mean molar mass compounds such as water are hard to drive off a planet in the quantities our delivery simulations suggest \citep{2018A&ARv..26....2L,2020SSRv..216..129O}. Similarly, the water mass fractions derived from our simulations are too large to be efficiently eroded by the drier impacts \citep{2020MNRAS.499.5334S,2020NatGe..13..265G}. The suggestion of a substantial influx of late volatiles from local and farther out debris shares some similarities with the hypothesis of atmospheric rejuvenation by exo-comets from extrasolar debris disks \citep{2018MNRAS.479.2649K,2020tnss.book..351W}. However, because rocky planets in M-dwarf systems likely undergo long-lived magma ocean phases -- as outlined above -- most of these volatiles will be dissolved in the magma ocean and stored in the mantle \citep{2021ApJ...922L...4D}. This lowers escape rates relative to the assumption that all volatiles are stored on the surface, and can potentially explain the origin of the density dichotomy of the TRAPPIST-1 system \citep{2019A&A...628A..12T,trappist_new} and the density distribution in the K2-3 system \citep{2022arXiv220712755D}.

\subsection{Caveats}

Our simulations explore the impact of varying physical devolatilization mechanisms in M-dwarf planetary systems by use of different scenarios that vary the water mass fraction, stellar mass, and planetesimal distribution. Water loss via photolysis from magma oceans, however, operates on timescales comparable to the pre-main sequence evolution of M-dwarfs. Therefore, a more refined treatment of magma ocean evolution in conjunction with planet formation and bombardment simulations is required in order to predict detailed compositions for individual systems -- as will be necessary for the exploration of individual planets in the JWST era. 

Our $N$-body models necessarily assume that all habitable zone planets around M-dwarfs formed via giant impacts in the same manner as the Solar System's terrestrial planets. Thus, our study does not incorporate a wide sweep of potential initial conditions that may arise if the role of pebble accretion is higher than in the Solar System. This may shift our estimates to lower impact rates, and we therefore regard our results as upper limit estimates on the rate of late impacts.  However, it is reasonable to argue that debris-producing giant impacts occur in most rocky-planet formation scenarios \citep[see further discussion in][]{clement21_mdwarf}, even if the majority of planet growth occurred via pebble accretion rather than direct planetesimal-planetesimal growth.  It is also important to note that a range of dynamical formation models currently purport to explain the origin of the inner Solar System \citep[e.g.:][]{ray20_rev}.  Worse still, different potentially viable evolutionary scenarios for the terrestrial planets yield disparate late bombardment chronologies \citep{clement19_frag}, and it remains challenging to constrain models in this manner by inferring an impact history from crater counts on terrestrial bodies \citep[e.g.:][]{evans18,brasser20_bombard}.  Thus, any attempt to neatly map Solar System science to the exoplanet regime should be viewed with a degree of skepticism.  Therefore, our study should be viewed more as an exploration of the range of plausible end-states, rather than a definitive model for any particular system.

Finally, throughout this work we use water mass fraction as a proxy for oxidation state. However, the impact of various loss mechanisms such as dehydration by short-lived radionuclides can fractionate the material redox state from the water mass fraction. For instance, this may happen via serpentinization processes on planetesimals during internal heating. Further work is required to quantify the redox evolution during the internal geophysical and geochemical evolution of planetesimals, both in the Solar System and in extrasolar planetary systems.

\subsection{Observational tests}

Our numerical experiments suggest that M-dwarf exoplanets -- and any surviving debris -- should be water-rich and oxidized. This is testable using astronomical observations of both individual exoplanets, and from statistical correlations via survey missions such as TESS and PLATO. On a statistical level, evolution is dominated by two effects: (i) desiccation via internal heating on planetesimals, and (ii) magma ocean losses on protoplanets. The former operates on a system level at early stages \citep{2019NatAs...3..307L,2021ApJ...913L..20L}, while the latter operates to different degrees over time and is sensitive to irradiation \citep{2016ApJ...829...63S,2022arXiv220310023L}.

If M-dwarf planetary systems in fact feature late impacts of reducing debris -- as is realised in a minor fraction of our simulations only -- then JWST may be able to observe the transient signature of this \citep{2020AsBio..20.1476F,2021ApJ...921L..28R}. Our simulations predict the observable time window for this to be in the first few tens of Myr across the M-dwarf mass spectrum. However, in such a case the observation would need to take into account the underlying magma ocean. Distinguishing between a redcued magma ocean scenario \citep{2021ApJ...914L...4L,2022PSJ.....3..127S} and an impact-induced reduced atmosphere \citep{2020ApJ...888...21R} may require observations in the near- to mid-infrared to differentiate the near-surface temperature \citep{2021JGRE..12606711L}. The dense sub-Earth GJ 367b \citep{2021Sci...374.1271L} may provide an attractive observational target, if its host star is within the first $\sim$100 Myr of its life, as suggested by \citet{2022MNRAS.513..661B}.

Finally, observational evidence from polluted white dwarfs suggests a fraction of the observed debris is core- or mantle material \citep[hence, the debris from differentiated planetesimals:][]{2020MNRAS.492.2683B}, while others demonstrate evidence for water-rich compositions \citep{2015MNRAS.450.2083R,2021ApJ...907L..35D}. Finding the source of these compositional trends will help to refine predictions for exoplanetary systems and the probability of reduced versus oxidized (and volatile-rich) late bombardment \citep{2020MNRAS.492.2683B,2022MNRAS.515..395C}.

\section{Summary \& Conclusions} \label{sec:conclusions}

We performed late-stage impact bombardment simulations with a focus on rocky exoplanets in the liquid water habitable zones of M stars between 0.1--0.4 $M_\odot$ with the goal of quantifying the chances of triggering transiently reducing conditions suitable for sub-aerial prebiotic chemistry in Hadean Earth analog environments. In order to test different initial planetary embryo and planetesimal compositions we sub-divided the disk into three regions; corresponding to inside (\emph{Region 1}) and outside (\emph{Region 2}) the steam runaway greenhouse limit, and outside the water snow line (\emph{Region 3}). Impact bombardment curves are derived based on two initial size frequency distributions, based off the modern asteroid belt and one including an additional contribution of planetesimals akin the outcome of the streaming instability. 

The compositions of planetary embryos and planetesimals in each bombardment scenario are initiated with a different water mass fraction, which we take as a first-order proxy for redox state. The water mass fraction of planetary embryos is assumed to be influenced by the orbit of the water snow line during the formation of the initial planetesimal swarm that formed the embryo, and water loss due to photolysis during primordial runaway greenhouse (magma ocean) episodes of the extended pre-luminous main sequence phase of M stars. Initial planetesimal compositions are assumed to be influenced by orbital location relative to the water snow line and dehydration based on internal heating by short-lived radioactive isotopes such as $^{26}$Al. Overall we compare five different general settings, varying the radial and temporal distribution of water mass fractions in embryos and planetesimals across the M-dwarf mass spectrum, which simulates different degrees of magma ocean desiccation, planetesimal dehydration, and planetesimal formation timescale relative to the radial evolution of the water snow line.

Summarising or results, we find:
\begin{itemize}
    \item The majority of bombardment epochs in our simulations are dominated by water-rich compositions, comparable to carbonaceous chondrite-like or cometary compositions, and hence oxidizing in bulk abundance, in contrast to the composition of the late veneer in the Solar System. Only the simulations that are set up to simulate the driest-possible end-member cases (\emph{Dry 3} and \emph{Desiccated Planetesimals 3}) achieve planetary bombardment episodes that include dry and hence reducing impacts that could conceivably trigger intermittent reducing climate states amenable for sub-aerial prebiotic chemistry. A late veneer-analog bombardment of chemically reduced planetesimals that deliver reducing power in the form of iron metal to react with vaporized oceans is thus only possible when planetesimals dehydrate substantially relative to their primordial, disk-derived compositions.
    \item Across the entire M-dwarf mass spectrum, late-stage bombardment of young rocky planets ceases before the star transitions onto the main sequence. Young exoplanets that incorporate \ce{H2O} or \ce{H2} abundances comparable to the early Earth -- as suggested by the density dichotomy between the super-Earth and sub-Neptune regimes and our planet formation simulations -- will be covered by global primordial magma oceans during this time window. Therefore, all late-stage debris will fall into the magma ocean without being able to chemically equilibrate with surface water.
    \item Integrated over all possible volatile delivery scenarios, we find the total bulk abundances of M-dwarf exoplanets to be strongly volatile-enriched compared to those of the Solar System's terrestrial planets. Based on previous estimates of atmospheric escape fluxes and the longevity of magma ocean episodes under strong irradiation, this suggests that M-dwarf exoplanets typically escape their primordial runaway greenhouse phases with sufficient bulk atmophile content to re-generate secondary atmospheres via outgassing from their mantle reservoir. From a compositional perspective, M-dwarf exoplanets with irradiation levels comparable to the modern Earth are not in danger of losing their atmospheres. Rather, the challenge is to get rid of an overabundance of volatiles to enable a potentially habitable surface.
    \item Solar-like planetary systems therefore are statistically more likely to experience intermittent reduced climate states early in their evolution -- a key requirement to initiate prebiotic synthesis in sub-aerial origin of life scenarios. In the absence of chemically reducing processes such as incomplete core-mantle differentiation in super-Earths, this suggests that M-dwarf rocky exoplanets might host oxidizing atmospheres and feature substantially decreased bulk densities relative to an Earth-like composition.
    \item In contrast, G-dwarf stars experience late-stage impacts for hundreds of millions of years after rocky planets in their liquid water habitable zones have solidified. In addition, G-dwarf planetary systems generally receive drier late-stage impacts than M-dwarf exoplanets. On average, the final bulk water abundances of G-star exoplanets are between one to two orders of magnitude lower compared to those of M-dwarf exoplanets.
\end{itemize}

Our results highlight the intimate connection between compositional variables set by early disk chemistry, geophysical internal processes, and dynamical planet formation scenarios that can affect the long-term atmospheric content and surface conditions of rocky exoplanets. Our simulations offer predictions that are testable via atmospheric and surface observations of short-period exoplanets in M-dwarf planetary systems, and compositional analyses of debris disks and polluted white dwarfs. Combining insights from the geophysical exploration of young, Hadean-analog exoplanets and the distribution of climate states across the rocky exoplanet census will enable a sharper picture of plausible prebiotic environments and pathways to the origin of life as we know it on our own world.
\vspace{-0.7cm}
\acknowledgments
The authors are grateful for useful discussions with Anat Shahar and Nick Wogan, as well as the constructive report of an anonymous reviewer. The work described in this manuscript was supported by the AEThER project, funded by the Alfred P. Sloan Foundation under grant No.~G202114194, the Simons Foundation (SCOL Award No.~611576), Carnegie Science's Scientific Computing Committee for High-Performance Computing (hpc.carnegiescience.edu), and benefitted from information exchange within the program ‘Alien Earths’ (NASA Grant No.~80NSSC21K0593) for NASA’s Nexus for Exoplanet System Science (NExSS) research coordination network. This work also used the Extreme Science and Engineering Discovery Environment (XSEDE), which is supported by National Science Foundation grant No.~ACI-1548562. Specifically, it used the Comet system at the San Diego Supercomputing Center (SDSC).  This research made use of resources provided by the Open Science Grid \citep{osg1,osg2}, which is supported by the National Science Foundation Award No.~1148698, and the U.S. Department of Energy's Office of Science. Some of the computing for this project was performed at the University of Oklahoma's Supercomputing Center for Education and Research (OSCER).

\newcommand{\noop}[1]{}

\end{document}